# Hierarchical Clustering Supported by Reciprocal Nearest Neighbors

Wen-Bo Xie, [1,2] Yan-Li Lee, [3] Cong Wang, [1] Duan-Bing Chen, [1,2,4] Tao Zhou[1,3]*

1 Big Data Research Center, University of Electronic Science and Technology of China, Chengdu 611731, People's Republic of China.
2 Union Big Data Tech. Inc., Chengdu 610041, People's Republic of China.
3 CompleX Lab, University of Electronic Science and Technology of China, Chengdu 611731, People's Republic of China.
4 The Center for Digitized Culture and Media, University of Electronic Science and Technology of China, Chengdu 611731, People's Republic of China.

**Abstract:** Clustering is a fundamental analysis tool aiming at classifying data points into groups based on their similarity or distance. It has found successful applications in all natural and social sciences, including biology, physics, economics, chemistry, astronomy, psychology, and so on. Among numerous existent algorithms, hierarchical clustering algorithms are of a particular advantage as they can provide results under different resolutions without any predetermined number of clusters and unfold the organization of resulted clusters. At the same time, they suffer a variety of drawbacks and thus are either time-consuming or inaccurate. We propose a novel hierarchical clustering approach on the basis of a simple hypothesis that two reciprocal nearest data points should be grouped in one cluster. Extensive tests on data sets across multiple domains show that our method is much faster and more accurate than the state-of-the-art benchmarks. We further extend our method to deal with the community detection problem in real networks, achieving remarkably better results in comparison with the well-known Girvan-Newman algorithm.

# Introduction

Clustering algorithm is of great importance in the studies of data mining. As an unsupervised machine learning method, it can help people to understand data without clearly preassigned labels. Therefore, clustering algorithm has already found many successful applications in disparate fields, such as biology, chemistry, physics and social science (*1*). Accordingly, scientists have been studying clustering algorithms for more than a half century, with a large number of excellent algorithms having been put forward and widely used. Well-known algorithms can be divided into different categories, such as the partition methods (*2*, *3*), the density-based algorithms (*4*, *5*), the affinity propagation algorithms (*6*, *7*), the feature transformation methods (*8-10*), and so on. The clustering results of these algorithms are often challenged by the poor readability, caused by the lack of observable representative data point for each cluster, the exogenously determined number of clusters together with the fixed granularity, and the unclear organization of resulted clusters.

Different from the above algorithms, a hierarchical clustering algorithm will produce a clustering tree, which clearly reflects the organization of resulted clusters and could provide clustering results under different resolutions without the help of a predetermined number of clusters (*11*, *12*). Such remarkable advantage of hierarchical clustering algorithms facilitates their applications in scientific analyses, including gene-related predictions (*13*, *14*), graph mining (*15*, *16*), human brain analyses (*17*, *18*), environmental assessment (*19*), incidence relation identification (*20*, *21*), and so on. Traditional hierarchical clustering algorithms are usually highly time-consuming, which suffer growing challenges from the increasing data volume in the so-called *big data era*.

Many fast algorithms for hierarchical clustering have been proposed, such as the algorithms via constructing the clustering feature tree (CF Tree) (*22*, *23*), the random sampling algorithms (*24*, *25*), the stepwise algorithms based on the nearest neighbor graph (*26*), and so forth. But these algorithms are subject to a common drawback: accompanied by the improvement of efficiency, the algorithms' accuracy drops. In a word, the following general problems are still to be optimized (*27-30*). (i) The scalability problem: the algorithm scales poorly in both memory and computing time with increasing data volume; (ii) The go-back problem: once two clusters merge into a new cluster, the new one cannot be unfasten; (iii) The chaining effect problem: a few data points (i.e. noisy data points) located between two weakly connected clusters may form a bridge so as to merge these two clusters into one, eventually resulting in highly skewed dendrograms.

Facing the above-mentioned challenges, this article proposes a novel hierarchical clustering algorithm (named as Reciprocal-nearest-neighbors Supported clustering, RS for short), which is based only on a compact hypothesis that the two reciprocal nearest data points should be put in one cluster. According to extensive experiments on University of California Irvine (UCI) data sets for machine learning (*31*) and Olivetti face data set (*32*), with very low computational complexity, RS algorithm provides more accurate results than classical benchmarks (e.g., group average (*33*) and CURE (*25*)) and state-of-the-art methods (e.g., affinity propagation (*6*) and clustering via density peaks (*4*)). We further extend RS algorithm to deal with the community detection problem (*16*, *34*) in networks and demonstrate its advantage in comparison with the well-known Girvan-Newman algorithm (*35*).

## Algorithm

The RS algorithm treats each data point as a node, and starts from an empty set of sub-clustering trees (SCTs for short) and a candidate set of all $n$ nodes. If the candidate set is not empty, one node therein, denoted by $i$, will be randomly selected. This node will be linked to its nearest neighbor $\delta_i^{(1)}$. Notice that, we assume the distance $d_{xy}$ between any two nodes $x$ and $y$ is well defined, and to avoid unnecessary complications caused by multiple nearest neighbors, we disturb each distance $d_{xy}$ by adding a very small random variable $\varepsilon_{xy} \ll d_{xy}$ as $d_{xy} \leftarrow d_{xy} + \varepsilon_{xy}$. After that, each node has only one nearest neighbor. $\delta_i^{(1)}$ will be further linked to its nearest neighbor, denoted by $\delta_i^{(2)}$. Setting $\delta_i^{(0)} = i$, such process will produce a chain $\delta_i^{(0)} \to \delta_i^{(1)} \to \cdots \to \delta_i^{(h)}$. The construction stops if one of the following two conditions holds: (i) $\delta_i^{(h)} = \delta_i^{(h-2)}$, which means $\delta_i^{(h-2)}$ and $\delta_i^{(h-1)}$ are reciprocal nearest neighbors (RNNs); (ii) $\delta_i^{(h)}$ is not in the candidate set. If the second condition holds, this chain will be linked to a certain existing SCT, while if the first condition holds, $\delta_i^{(0)} \to \delta_i^{(1)} \to \cdots \to \delta_i^{(h-1)}$ will constitute a new chain-shaped SCT, with $\delta_i^{(h-2)}$ and $\delta_i^{(h-1)}$ being the two supporting nodes. A unique artificial root, as the representative of this SCT, will be linked to these two supporting nodes. After that, all nodes $\delta_i^{(0)}, \delta_i^{(1)}, \cdots, \delta_i^{(h-1)}$ will be removed from the candidate set. If the candidate set is not empty, one node therein will be randomly selected and the above process will be implemented again. Otherwise the whole procedure to construct SCTs will stop.

The next task is to prune the SCTs to avoid some very long and thin SCTs, which may accumulate transmitted errors and thus reduce the algorithmic performance. For any node $i$, if it belongs to an SCT with two supporting nodes $p$ and $q$, its depth is defined as

$$h_i = \frac{1}{2}\left(l_{ip} + l_{iq} + 1\right), \tag{1}$$

where $l_{ip}$ is the shortest path length between $i$ and $p$ in this SCT. Though with an artificial root, $p$ and $q$ are assumed to be directly connected, so that $|l_{ip} - l_{iq}| = 1$. Therefore, $h_i$ can be considered as the path length between $i$ and the artificial root. For each SCT $C$, all nodes whose depths are larger than a threshold $\Phi(C)$ will be pruned. Every pruned node will be linked to an artificial root and form a special SCT consisted of only one root and one node. The threshold is defined as

$$\Phi(C) = \lceil \log_\alpha(|C|+1) \rceil, \tag{2}$$

where $|C|$ is the number of nodes in $C$, and $\lceil x \rceil$ denotes the smallest integer no smaller than $x$. In later experimental tests, the value of $\alpha$ is fixed as $\alpha = 1.5$ and the sensitivity analysis on the selection of the parameter $\alpha$ are presented in Figure S1 in the Supplementary Materials.

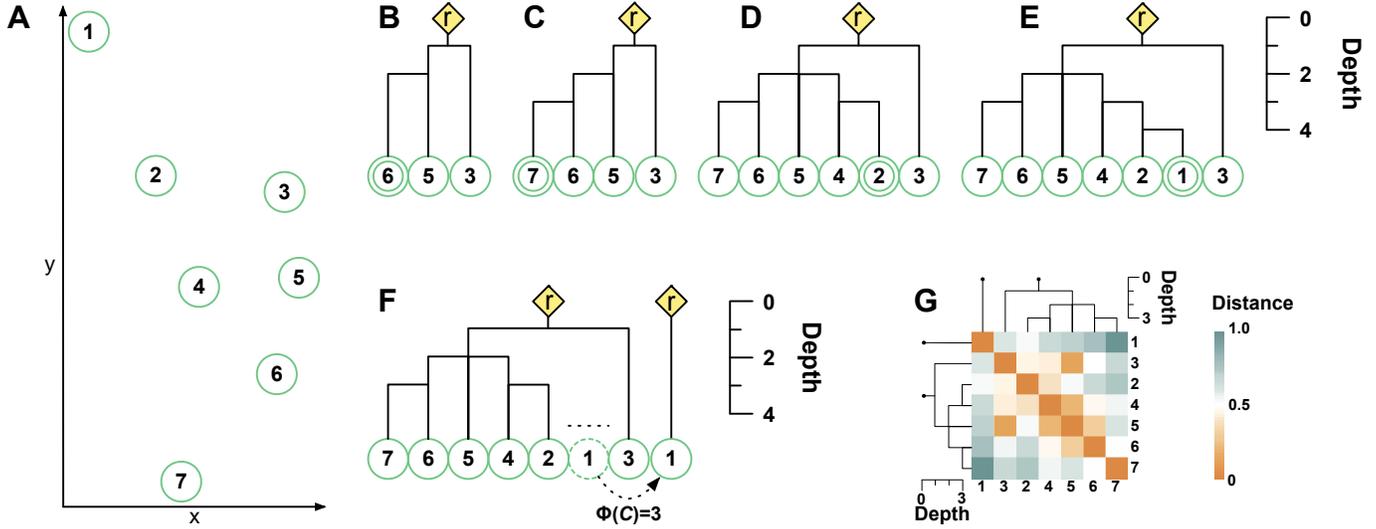

**Fig. 1. Illustration of the procedure to construct and prune the sub-clustering trees.** (**A**) The original distribution of the seven data points in a two-dimensional plane. (**B**) Node 6 is randomly selected as the initial node to form the chain that ends with nodes 3 and 5 being the two RNNs. An artificial root is thus assigned to the two supporting nodes. (**C**) Node 7 is randomly selected from the updated candidate set {1,2,4,7}, whose nearest neighbor is node 6, so it is directed linked to node 6. (**D**) Node 2 is randomly selected from the updated candidate set {1,2,4}, whose nearest neighbor is node 4, and node 4's nearest neighbor is node 5. Hence this chain will be linked to node 5 as $2 \to 4 \to 5$. (**E**) The last node in the candidate set, node 1, will be linked to its nearest neighbor, say node 2. Plot (**E**) illustrates the full structure of the SCT with nodes 3 and 5 being the two supporting nodes. Assuming $\alpha = 2$, so $\Phi(C) = 3$, and thus node 1 will be pruned since $l_1 = 4$. Node 1 will form a special SCT as shown in plot (**F**). Plot (**G**) shows the heatmap of the normalized distance matrix as well as the structure of SCTs after pruning, where the normalized distance between any two nodes $i$ and $j$ is defined as $d_{ij}/d_{max}$, in which $d_{ij}$ is the Euclidean distance between nodes $i$ and $j$, and $d_{max}$ is the maximum Euclidean distance between any pair of nodes.

Figure 1 illustrates a simple example where 7 data points are distributed in a two-dimensional plane, and the distance between any two nodes is defined as their Euclidean distance. After the construction of SCTs, node 1 is pruned and thus there are eventually two SCTs. Generally speaking, there are many SCTs after pruning. Then, we treat each SCT as a node (represented by its root) and implement the above constructing and pruning processes again to obtain a higher-level clustering. Such procedure will be implemented iteratively, resulting in the final clustering tree. Initially, all roots are put in the candidate set. The location of a root is either identical to its linked node if it represents a special SCT with only one node being pruned in the last pruning process, or defined as the midpoint of the two supporting nodes it links to. The distance between two roots is thus defined as the Euclidean distance between their locations. From the initial data distribution to the final result, a full procedure of the proposed RS algorithm is illustrated in Figure 2.

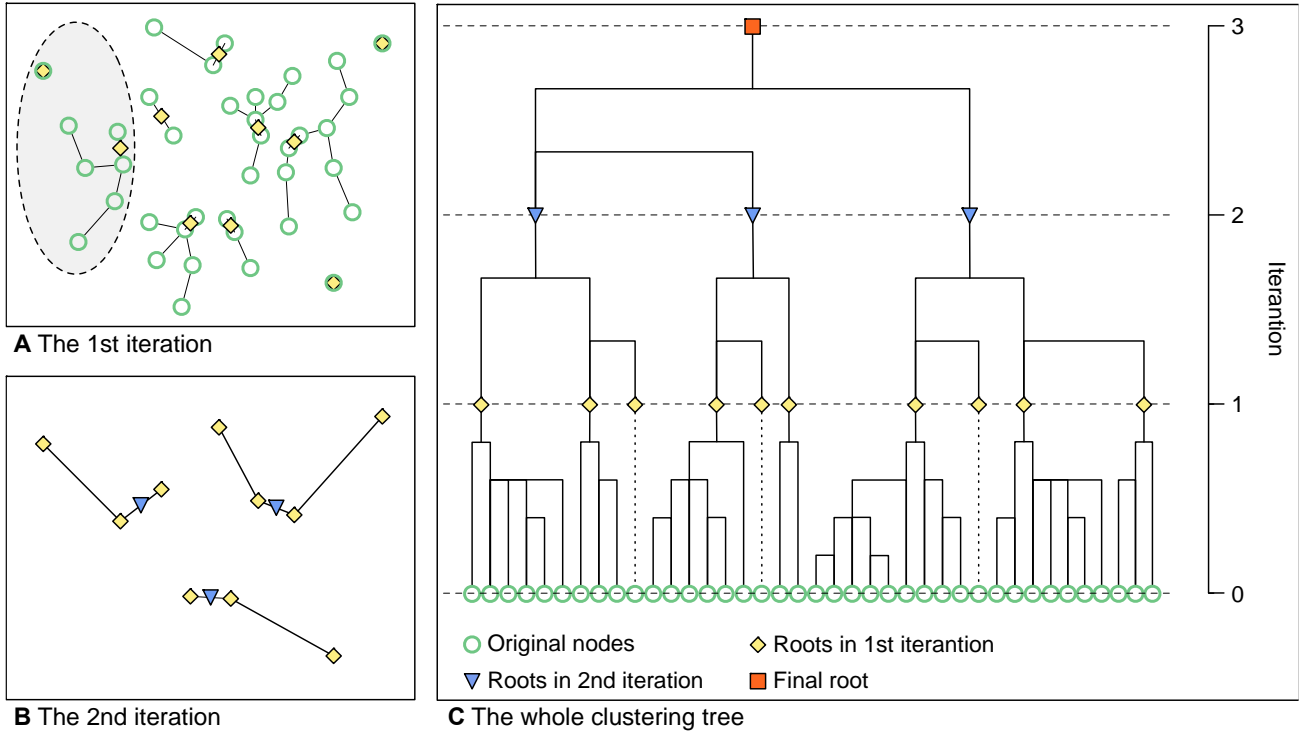

**A** The 1st iteration
**B** The 2nd iteration
**C** The whole clustering tree

**Fig. 2. Illustration of the full procedure of the RS algorithm.** (**A**) The initial data points (green circles) and the result of the first iteration where roots are represented by yellow diamonds. Notice that, the SCTs emphasized by shadow are the same to those shown in Figure 1. (**B**) The result of the second iteration where the generated roots in this iteration are represented by blue triangles. (**C**) Final clustering tree obtained after three iterations, where the red square is the highest-level root.

In a more general case where data points cannot be properly embedded in an Euclidean space and only a distance between each pair of data points is well defined (see the last two examples in the next section), we can still construct the SCTs but the distance between two roots cannot be updated as mentioned above, because the specific locations of roots are not well defined. In such circumstance, the distance between two roots is updated using only the information of distances instead of locations. Denoting $r_1$ and $r_2$ the two roots under consideration, via some simple geometrical calculation, the distance between $r_1$ and $r_2$, say $|r_1 r_2|$, can be determined according to the following three different situations: (i) if $r_1$ is identical to a single point $a$, and $r_2$ is identical to a single point $c$, then $|r_1 r_2| = |ac|$; (ii) if $r_1$ is the midpoint of $a$ and $b$, and $r_2$ is identical to a single point $c$, then $|r_1 r_2| = \sqrt{\frac{|ac|^2 + |bc|^2}{2} - \frac{|ab|^2}{4}}$; (iii) if $r_1$ is the midpoint of $a$ and $b$, and $r_2$ is the midpoint of $c$ and $d$, then $|r_1 r_2| = \frac{1}{2}\sqrt{|ac|^2 + |ad|^2 + |bc|^2 + |bd|^2 - |ab|^2 - |cd|^2}$.

## Results

We test the performance of the proposed RS algorithm on eight selected data sets from the UCI database (*31*), which is well recognized as a standard database for machine learning. The detailed information about these data sets is presented in the Materials and Methods. Two classical clustering algorithms (i.e., the group average (GA) (*33*) and CURE (*25*)) and two state-of-the-art methods (i.e., affinity propagation (AP) (*6*) and clustering via density peaks (DP) (*4*)) are used for comparison. Detailed information about these benchmarks is shown in the Materials and Methods. The Rand Index (*36*) is adopted to quantify the algorithms' performance: the larger the index is, the better the clustering result is. The mathematical definition of the Rand Index is presented in the Materials and Methods.

Figure 3 compares the performance of RS and other considered algorithms on the UCI data sets. It can be observed that RS is competitive with AP and performs overall better than other algorithms. In particular, for mfeat-fourier and optidigits, many algorithms perform poorly while RS shows decent clustering ability. In addition, we examine the efficiency of RS by comparing the required CPU time of RS with other algorithms. As shown in Figure S2 in the Supplementary Materials, RS is the fastest one among all considered algorithms, while GA and AP are of the highest and the second highest time complexity.

We next test the RS algorithm on the Olivetti face data set (*32*), which is a benchmark data set for clustering and classification. This data set consists of 400 real face images in 40 categories, with each category containing a person's 10 different face images. Figure 4(**A**) illustrates the resulted clustering tree for the first 50 face images by RS. As shown in Figure 4(**B**), overall speaking, RS performs best. In particular, after two iterations, RS obtains 39 clusters (very close to the value $K=40$ for the ground truth) with Rand Index being 0.976. The full illustration is shown in Figure S3 in the Supplementary Materials.

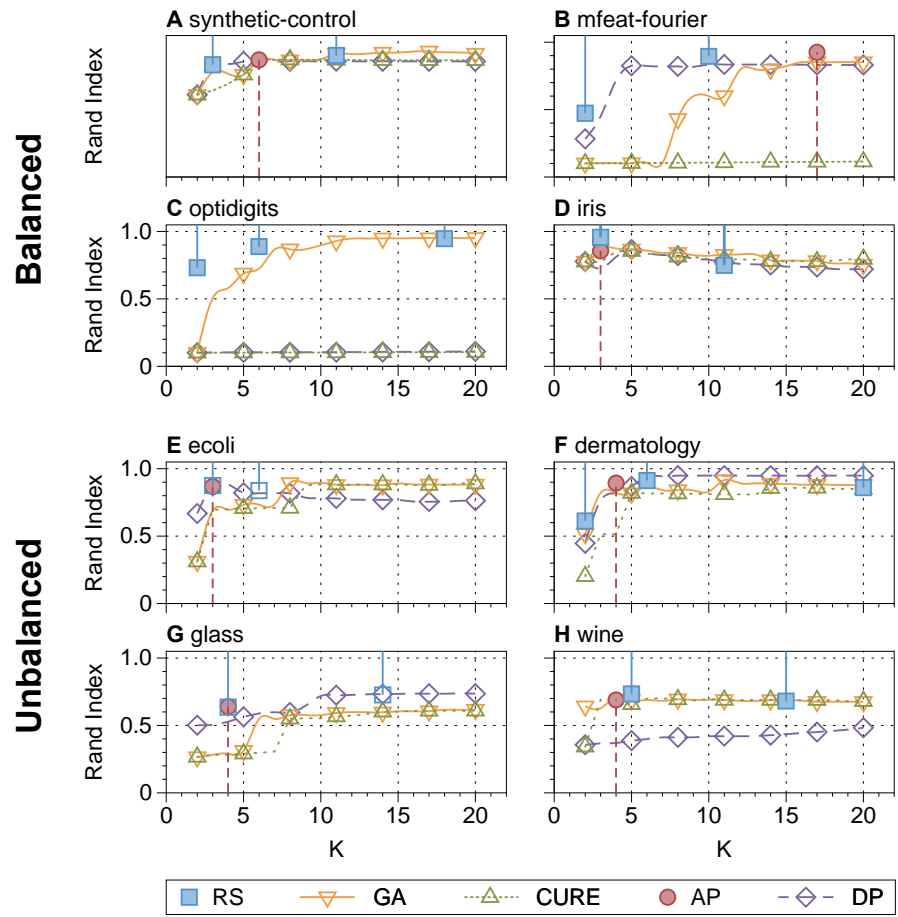

**Fig. 3. Performance of algorithms on the UCI data sets.** The X-axis denotes the number of clusters $K$, the Y-axis shows the Rand Index, and each plot corresponds a selected data set. For GA, CURE and DP, $K$ is preassigned and thus adjustable, while for RS and AP, $K$ is endogenous. For example, in RS, different iterations correspond to different values of $K$. To be clear to readers, the results of RS and AP are respectively emphasized by vertical blue solid lines and vertical red dash lines. The number of cluster, $K=38$, produced by AP on optidigits is much larger than the value of the ground truth (i.e., $K=10$), so that we don't show the result of AP in plot (**C**).

The RS algorithm can also be extended to detect communities, which is a significant long-standing challenge in network science (*16*, *34*). The community structure is loosely defined as a number of communities where each node belongs to one community and connections within a community are much denser than connections in between communities (*35*). The community detection problem can be considered as a specialized clustering algorithm where nodes are treated as data points (*34*), and thus a necessary step before transforming community detection problem to clustering problem is to determine the similarity or distance between node pairs (*38*).

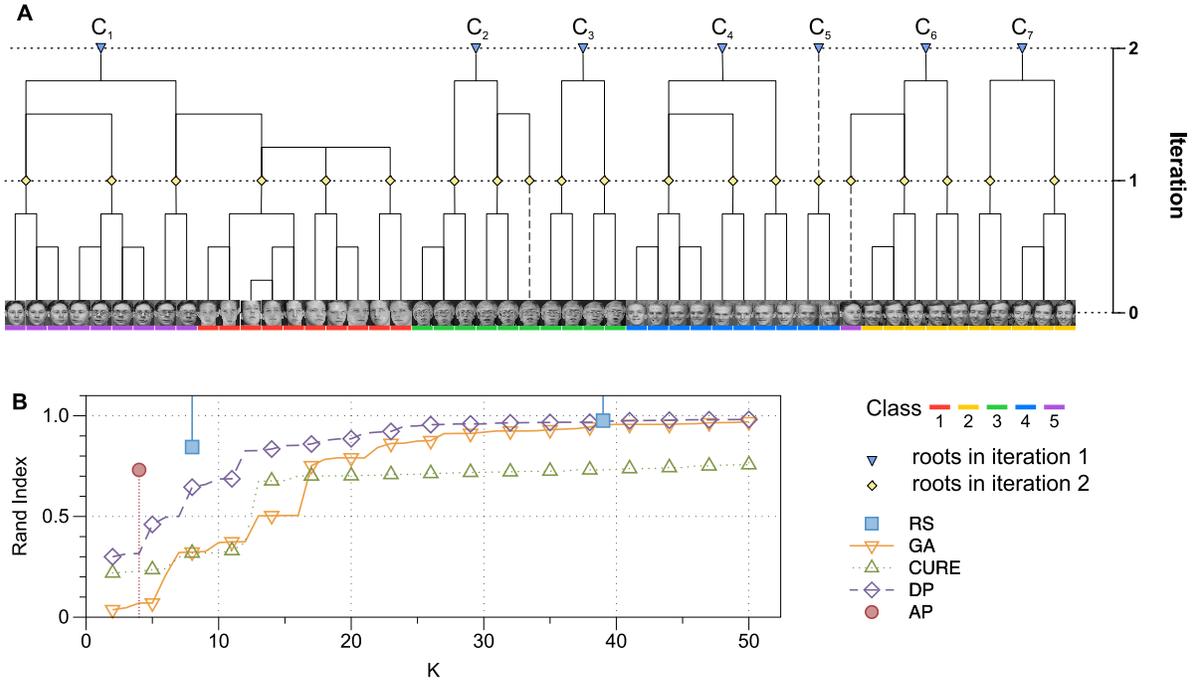

**Fig. 4. Result and Performance of algorithms on the Olivetti face data set.** Plot (**A**) illustrates the resulted clustering tree for the first 5 persons' faces (i.e., 50 images) by RS, where images in the same color belong to one category in the ground truth. Plot (**B**) compares the performance of RS with other considered algorithms. In all relevant algorithms, the similarity between two face images is determined by the complex wavelet structural similarity (*37*), and the corresponding distance is defined as the inverse of the similarity.

Denote $A$ the adjacency matrix of an undirected network $G$. For the unweighted case, $a_{ij} = 1$ if nodes $i$ and $j$ are adjacent (being directed connected by an edge), otherwise $a_{ij} = 0$. For the weighted case, $a_{ij} = 0$ if nodes $i$ and $j$ are not adjacent, otherwise $a_{ij}$ represents the weight of the edge $(i, j)$. The Laplace matrix of $G$ is then defined as $L = D - A$, where $D$ is a diagonal matrix with its diagonal element $d_{ii}$ being equal to the degree of node $i$. We apply the random walk theory (*39*) to define the distance between any two nodes as

$$D(i, j) = \sqrt{V_G(l_{ii}^+ + l_{jj}^+ - 2l_{ij}^+)}, \qquad (3)$$

where $V_G = \sum_{i=1}^{N} d_{ii}$, $N$ is the number of nodes in $G$, and $l_{\bullet\bullet}^+$ is the corresponding element of the Moore-Penrose inverse matrix of $L$ (*40*). In the extended RS algorithm, each node is treated as a data point, the distance between two adjacent nodes is calculated by Eq. (3), and the distance between two disadjacent nodes is set as infinite. Due to the sparsity of the network, the algorithm will stop when the distance between any two roots is infinite. At that time, there may be some isolated nodes resulted from the pruning process, as well as some isolated reciprocal nearest neighbors. Each of these nodes, no matter a single node or one of RNNs, will be merged to the its nearest community whose size, right after the stop of algorithm, is no less than 3. Here the distance between a node $i$ and a community $C$ is defined as $D(i, C) = \min_{j \in C} D(i, j)$.

We test this extended RS algorithm on four real networks: beach (*41*), netsci (*42*), jazz (*43*) and haverford (*44*), of which two are unweighted and other two are weighted (see Materials and Methods for detailed description). The corresponding numbers of communities obtained by RS for the above four networks are 6, 30, 24 and 124, respectively. Statistically speaking, the sizes of communities are relatively small. Therefore, to overcome the possible resolution problem in community detection (*45*), we scan all edges in between communities, then at each time step we select the edge with the smallest betweenness (46) (see the definition in Materials and Methods) and merge the two communities bridged by this edge. This operation allows us to obtain divisions with fewer communities. We use the triangle participation ratio (47) (TPR for short, see the definition in Materials and Methods) as the metric to evaluate the algorithms' performance. As shown in Fig. 5, in comparison with the well-known Girvan-Newman algorithm (35) (see Materials and Methods for the introduction), in most cases, the extended RS algorithm performs remarkably better.

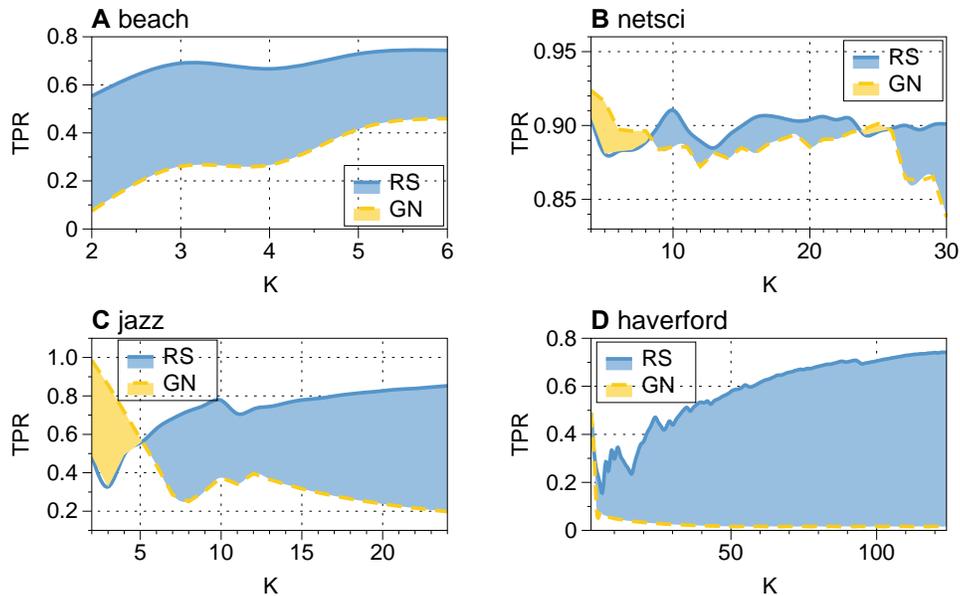

**Fig. 5. Performance of the RS and GN algorithms in community detection.** Each plot presents the result of a real network, with X-axis denoting the number of communities $K$ and Y-axis showing the metric of performance (i.e., TPR). The TPRs for RS and GN are blue solid curves and yellow dash curves, respectively. The blue shadows emphasize the regions where RS outperforms GN, while the yellow shadows indicate the opposite cases.

## Discussion

On the basis of a very intuitive hypothesis that two reciprocal nearest data points should belong to the same cluster, we propose a simple clustering algorithm that outperforms related techniques on both efficiency and accuracy. Beyond the better performance, our method has a strong interpretation power since it provides a clustering tree that records the hierarchical organization of resulted clusters under different resolutions. We further devise a way to calculate the distance between two artificial roots when their locations cannot be explicitly determined, so that we can extend our method to deal with some related yet different challenges, like the community detection problem in network science. Because of its simplicity, extendibility, efficiency and effectivity, we believe our method will find wide applications in natural science, social science, engineering, and so on.

To keep a record of artificial roots and to determine the distance between roots respectively ask for additional space and time. Therefore, if one can quickly evaluate which one of the two reciprocal nearest neighbors is the better representative data point, then the artificial roots are not necessarily introduced and the algorithm's efficiency can be further improved. We leave this idea as an open issue for future studies. Considering the extended algorithm for community detection, the currently applied method (39) to determine the distance between two nodes is highly time-consuming. Hopefully we can find out known measures or design novel measures on similarity or distance between nodes (see examples in the review article (49) and references therein) that can produce equally good or even better results in a much shorter time.

# Materials and Methods

## Data Description

**UCI database** (*31*) is one of the most well-known standard scientific databases, which maintains more than 400 data sets as a service to the machine learning community and is continuously updated. Eight labeled data sets are selected to test the performance of the proposed RS algorithm. Four of them are balanced data sets (i.e., each category contains approximately equal number of data points) and the others are unbalanced data sets. Parameters of these data sets are as shown in the Table 1.

**Table 1**: **Basic information of the eight selected UCI data sets.**

| Type | Data Set | #Samples | #Features | #Classes |
|---|---|---|---|---|
| Balanced | synthetic-control | 600 | 60 | 6 |
| | mfeat-fourier | 2000 | 76 | 10 |
| | optdigits | 5630 | 64 | 10 |
| | iris | 600 | 4 | 3 |
| Unbalanced | ecoli | 336 | 7 | 8 |
| | dermatology | 366 | 34 | 6 |
| | glass | 214 | 9 | 2 |
| | wine | 178 | 13 | 3 |

**Networks.** Four real networks are selected to test the performance of the RS algorithm on community detection. (i) **beach** (*41*) is a human contact network of windsurfers. Each node represents a sportsman, and the weight of an edge represents the number of contacts. (ii) **netsci** (*42*) is the largest connected component of the co-authorship network of scientists in network science. The weight of an edge denotes the number of corresponding co-authorized papers. (iii) **jazz** (*43*) is a collaboration network among jazz musicians, where two musicians are connected if they have played together in a band. (iv) **haverford** (*44*) is a social friendship network extracted from Facebook.

**Table 2**: **Basic information of the four networks.**

| Type | Name | #Nodes | #Edges | Average Degree | Clustering Coefficient |
|---|---|---|---|---|---|
| Weighted | beach | 43 | 336 | 15.63 | 0.564 |
| | netsci | 379 | 913 | 4.82 | 0.368 |
| Unweighted | jazz | 198 | 2742 | 27.70 | 0.520 |
| | haverford | 1446 | 59589 | 82 | 0.323 |

## Benchmark Clustering Methods

**GA** (*33*) is a typical hierarchical clustering algorithm which agglomerates sub-clusters according to their distances. That is, two sub-clusters that have the minimum distance will be merged. GA is also named as average-linkage method since the distance between two clusters in this algorithm is defined as the mean distance between elements of each cluster.

**CURE** (*25*) is an efficient hierarchical algorithm, which introduces a sampling strategy before agglomeration. Two main stages of CURE are as follows. (i) Choosing a constant number of well scattered nodes of a cluster and shrinking them towards the centroid of the cluster. These shrunk nodes are treated as the representatives of the cluster. (ii) Agglomerating sub-clusters according to the criteria like GA.

**AP** (*6*) is a novel clustering algorithm based on the message passing between nodes. Central nodes of clusters are recognized by alternating two message passing steps, in which two matrices, responsibility and availability, are updated iteratively. The process will stop after a preassigned number of iterations, or when the elements in clusters remain unchanged for 10 iterations. Those nodes whose 'responsibility plus availability' is positive are marked as the central nodes of clusters.

**DP** (*4*) is a density-based clustering algorithm, in which each node has two quantities, its local density and its minimum distance from nodes of higher density, which are used to recognize the central nodes of clusters. Each node is assigned to the same label as its nearest neighbor of higher density.

## *Rand Index*

The Rand Index (*36*) is a widely used evaluation index for clustering algorithms, which measures the similarity between two data partitions. Given a set $S$ of $n$ data points and its two partitions, the real partition $X$ and the algorithm-produced partition $Y$, the Rand Index $R$ is defined as

$$R = \frac{a+b}{a+b+c+d}, \tag{4}$$

where $a$ is the number of pairs in $S$ that are in the same subset in $X$ and in the same subset in $Y$; $b$ is the number of pairs in $S$ that are in the different subsets in $X$ and in the different subsets in $Y$; $c$ is the number of pairs in $S$ that are in the same subset in $X$ but in the different subsets in $Y$; $d$ is the number of pairs in $S$ that are in the different subsets in $X$ but in the same subset in $Y$.

## *Betweenness Centrality*

Betweenness Centrality (*46*) of a node $i$ is defined as

$$BC(i) = \sum_{s \neq i, s \neq t, i \neq t} \frac{g(s,i,t)}{g(s,t)}, \tag{5}$$

where $g(s,t)$ is the number of shortest paths between $s$ and $t$, and $g(s,i,t)$ is the number of shortest paths between $s$ and $t$ that pass through node $i$.

## *Triangle Participation Ratio*

Triangle Participation Ratio (TPR) (*47*) is defined as the ratio of nodes in a community that belong to the triadic closure embedded in the corresponding community. The TPR indicator of the community $C$ is defined as follows:

$$f(C) = \frac{|\{u : u, v, w \in C, \ a_{uv} a_{uw} a_{vw} = 1\}|}{n_C}, \tag{6}$$

where $n_C$ denotes the number of nodes in the community $C$, and $a_{uv} = 1$ if there is an edge between nodes $u$ and $v$, otherwise, $a_{uv} = 0$. The TPR value of one network is $\frac{1}{|\mathbb{C}|} \sum_{C \in \mathbb{C}} f(C)$, where $\mathbb{C}$ is the sets of communities by a certain method.

## *Girven-Newman Algorithm*

The Girven-Newman (GN) algorithm (*35*) is a classical hierarchical community detection algorithm including the following steps: (i) calculating the betweenness centrality of each edge; (ii) removing the edge with the largest betweenness centrality; (iii) calculating the betweenness centralities for all remaining edges; (iv) repeating the steps of (ii) and (iii) to obtain a series of community divisions.

## Acknowledgments

**Funding:** This work is partially supported by the National Natural Science Foundation of China under Grant Nos. 61433014 and 61673085, by the Science Strength Promotion Program of the UESTC under Grant No. Y03111023901014006, and by the Fundamental Research Funds for the Central Universities under Grant No. ZYGX2016J196.

**Author contributions:** W.B.X., D.B.C. and T.Z. conceived the research, W.B.X., Y.L.L. and C.W. performed the experiments, W.B.X, Y.L.L., C.W., D.B.C. and T.Z. analyzed the data. T.Z. wrote the manuscript.

**Competing interests:** The authors declare that they have no competing interests.

**Data and materials availability:** All the data used in the experiments are available online. UCI data sets are available in [http://archive.ics.uci.edu/ml], Olivetti face data set is available in [http://www.cl.cam.ac.uk/research/dtg/attarchive/facedatabase.html], and networks are available in [http://networkrepository.com].

# Supplementary Materials

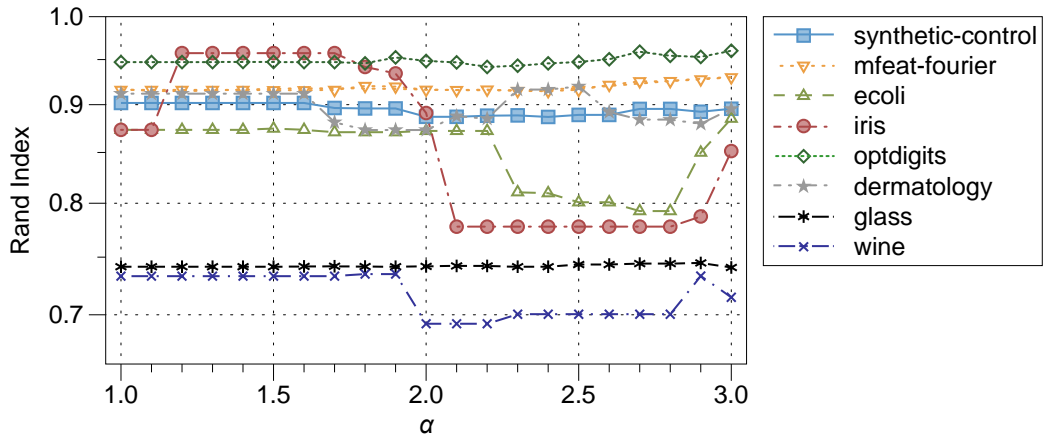

**Fig. S1.** Sensitivity analysis on the value of the parameter $\alpha$ for the eight UCI data sets. Each point denotes the best Rand Index in iterations with the corresponding $\alpha$. The algorithm's performance for synthetic-control, mfeat-fourier, optdigits and glass is not sensitive to $\alpha$, while the performance for ecoli, iris, dermatology and wine is relatively more sensitive to $\alpha$. Fortunately, for almost all cases, $\alpha = 1.5$ will produce the best or nearly the best results subject to the Rand Index, and thus in this paper, we fix $\alpha = 1.5$. If we are allowed to freely tune the parameter $\alpha$, we can achieve slightly better performance than that reported in Fig. 3, Fig. 4 and Fig. 5. However, it is not fair to other benchmarks, so that we only report the results corresponding to the fixed $\alpha$.

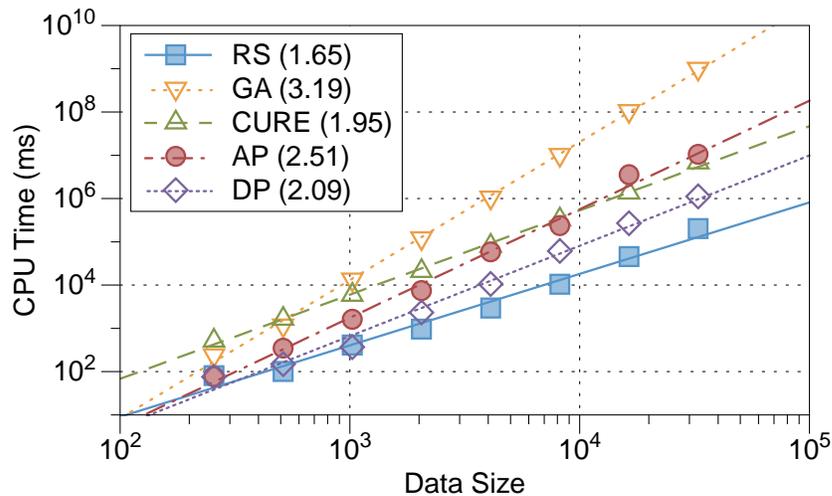

**Fig. S2.** CPU times on random data sets. We compare the required CPU times of RS algorithm with other algorithms on randomly generated data sets with different sizes. The relation between CPU times and data sizes for each algorithm is fitted by a power-law curve, with the exponent shown in the legend. Obviously, RS is the fastest algorithm while GA and AP are the slowest ones. The computation is implemented via a 3.4GHz 4 core Intel Core i5 CPU.

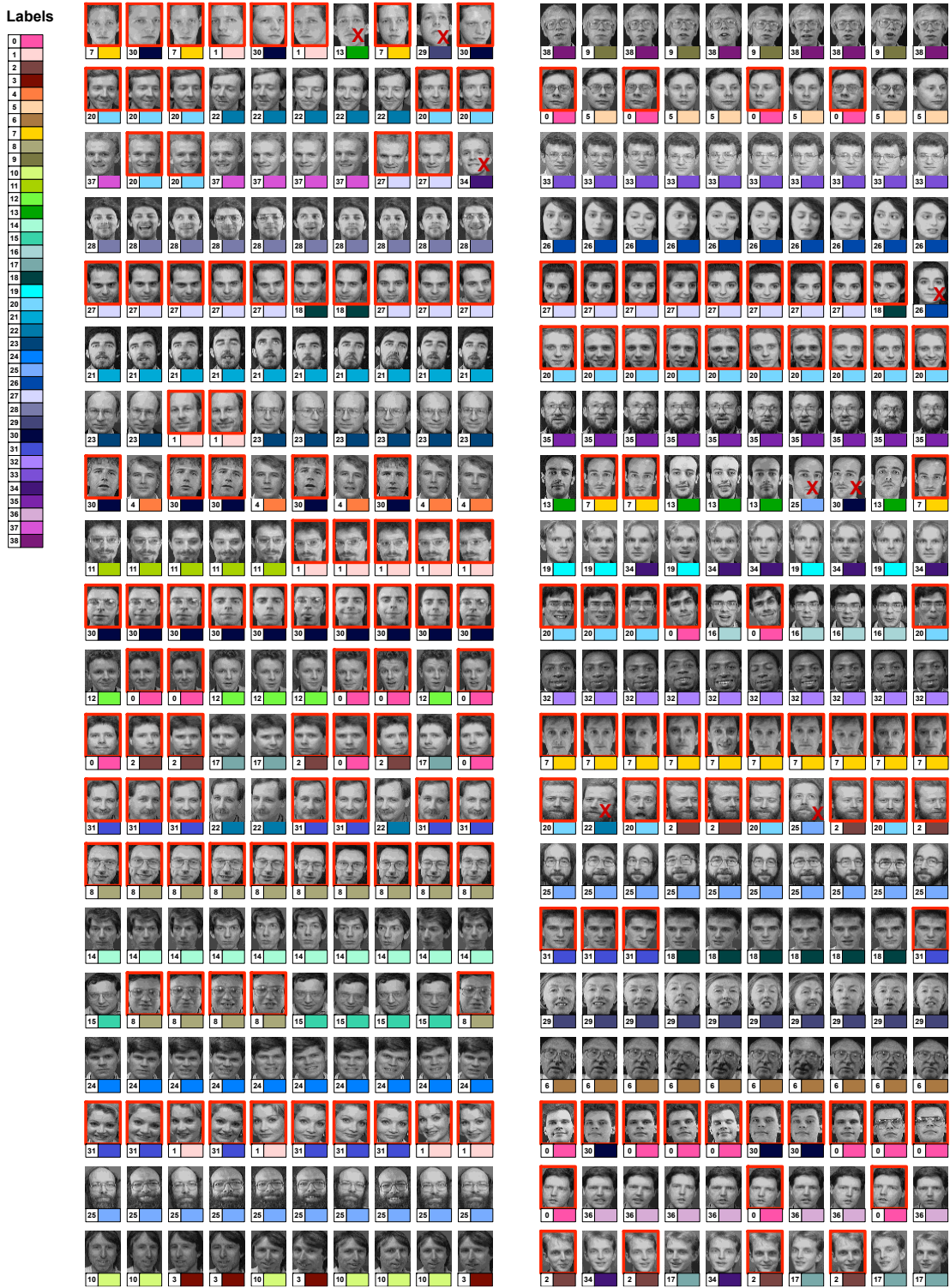

**Fig. S3.** Result of RS on the whole Olivetti face data set. As shown in this figure, after two iterations we obtain the result, where face images that belong to the same cluster are marked in the same number and color. Notice that, there are several impure clusters which contain multiple person's face images, which are marked by red boxes. In addition, red crosses are used to mark those incorrectly classified face images.